\documentclass[english,pre,showpacs,floatfix,preprint]
{revtex4}
\pdfoutput=1
\usepackage[T1]{fontenc}
\usepackage[latin1]{inputenc}
\usepackage{babel}
\usepackage{epstopdf}
\usepackage{amsmath}
\usepackage{amssymb}
\usepackage{graphicx}

\begin{document}

\title
{\bf Directed transport of two interacting particles in a washboard
potential}
\author{D. Hennig}
\author{A. D. Burbanks}
\author{A. H. Osbaldestin}
\medskip
\medskip
\medskip
\affiliation{Department of Mathematics, University of Portsmouth, Portsmouth, PO1 3HF, UK}

\begin{abstract}
\noindent We study the conservative and deterministic dynamics of two
nonlinearly interacting particles evolving in a one-dimensional
spatially periodic washboard potential. A weak tilt of the washboard
potential is applied biasing one direction for particle transport.
However, the tilt vanishes asymptotically in the direction of
bias. Moreover, the total energy content is not enough for both
particles to be able to escape simultaneously from an initial
potential well; to achieve transport the coupled particles need to
interact cooperatively.  For low coupling strength the two particles
remain trapped inside the starting potential well permanently.  For
increased coupling strength there exists a regime in which one of the
particles transfers the majority of its energy to the other one, as a
consequence of which the latter escapes from the potential well and
the bond between them breaks.  Finally, for suitably large couplings,
coordinated energy exchange between the particles allows them to
achieve escapes --- one particle followed by the other --- from
consecutive potential wells resulting in directed collective motion.
The key mechanism of transport rectification is based on the
asymptotically vanishing tilt causing a symmetry breaking of the
non-chaotic fraction of the dynamics in the mixed phase space.  That
is, after a chaotic transient, only at one of the boundaries of the
chaotic layer do resonance islands appear.  The settling of
trajectories in the ballistic channels associated with transporting
islands provides long-range directed transport dynamics of the
escaping dimer.

\end{abstract}

\pacs{05.60.Cd, 05.45.Ac, 05.60.-k, 05.45.Pq}{}\maketitle

\section{Introduction}

\noindent The study of transport phenomena has attracted considerable
interest over the years due to its relevance in many physical
situations. The latter are often described on the basis of
one-dimensional particle motion in a tilted spatially periodic
potential \cite{Faucheux}-\cite{Reimann}. Corresponding experimental
realisations include Josephson junctions \cite{Josephson}, charge
density waves \cite{charge}, superionic conductors \cite{superionic},
rotation of dipoles in external fields \cite{dipoles}, phase-locked
loops \cite{loops} and diffusion of dimers on surfaces
\cite{surfaces,Braun,Pijper,Braun1,Heinsalu,Patriarca,Fusco,Goncalves}
to name but a few. In many of these aforementioned situations the
particles, in addition to their motion in the periodic potential,
interact, which may lead to cooperative effects not found in
situations of individual particle motion
\cite{Evstigneev}-\cite{coop2}.

The objective of the current work is to investigate the conditions
under which it is possible to generate a directed flow along with
collective motion in a system of coupled particles. To be precise, we
study the transport of a dimer evolving in a washboard potential
experiencing a weak tilt force. The nonlinear bond dynamics between
the two monomers, constituting the dimer, is modelled by a Morse
potential allowing for bond breaking, i.e.  fragmentation. We focus
our interest on the chaos-promoted detrapping mechanism for dimers
that initially reside in one well of the washboard potential. Provided
that such a detrapping transition happens the question then is under
which circumstances subsequent directed long-range particle transport
is achievable.  Since the total system energy is too low for both
monomers to be able to escape from the potential well simultaneously,
we explore whether cooperative energy redistribution is possible
allowing at least one of the monomers to escape and subsequently
display directed motion. We also elucidate the possible scenario in
which the energy exchange between the monomers proceeds in such a
well-coordinated manner that the monomers move separately from one
well into the next, one following the other, resulting in directed
motion of the dimer.

The paper is organised as follows: In the next section the model of
the dimer system is introduced, followed by the formulation of the
escape problem together with a brief discussion of the related phase
space structure. In Section \ref{section:current} the particle current
is studied and the occurrence of different transport scenarios is
described.  Afterwards in Section \ref{section:space} we relate the
phase space dynamics to the regime of high particle current. In
particular chaotic invariant sets, their connection with singularities
of the escape time function, and their relevance for the escape process
are considered. In Section \ref{section:potential} we present an
alternative description of the escape problem as the motion of a
single particle in a two-dimensional potential landscape. Finally we
summarise and discuss our results.

\section{The model of the dimer system}\label{section:model}

\noindent We study the dimer dynamics with a Hamiltonian of the following form
\begin{eqnarray}
H&=&\sum_{n=1}^{2}\left[\frac{p_n^{2}}{2}+U_0(q_n)+U_1(q_n)\right]
+H_{int}(q_1,q_2)
\,,\label{eq:Hamiltonian}
\end{eqnarray}
wherein $p_n$ and $q_n$, $n=1,2$,  denote the canonically conjugate momenta
and positions of the two coupled particles 
of unit mass
evolving in the periodic,
spatially-symmetric washboard potential of unit period given by
\begin{equation}
U_0(q)=U_0(q+1)=-\,\frac{\cos(2\pi q)}{2\pi}\,.\label{eq:U0}
\end{equation}
The external field
\begin{equation}
U_1(q)=-F(q-\log[\cosh(q-q_0)])\label{equation:tilt}
\end{equation}
exerts a tilt on the washboard potential. The potential is sketched in
Fig.~$1$ for tilt strength $F=0.01$.  The value of the parameter
$q_0=10$ in the second term on the right hand side of
Eq.~(\ref{equation:tilt}) is chosen such that the tilt rapidly
diminishes when the coordinate $q$ exceeds $q_0$ and eventually upon
further growth of $q$ the bias vanishes.  On the other hand as long as
$q\ll q_0$ the tilt adopts the value $2F$. Therefore particles that
manage to escape from a potential well into the {\it asymptotic
region} $q_0<q\rightarrow \infty$ experience only a finite
acceleration period at the end of which any forward motion must
proceed unbiased. The question then arises whether escaping particles
carry on moving forward even in the range where the bias is no longer
present.

The interaction part of the Hamiltonian $H_{int}$ given by
\begin{equation}
H_{int}=\frac{D}{2}\left(1-\exp[-\alpha(q_{2}-q_{1}-l_0)]\right)^2\,,
\label{equation:Hint}
\end{equation}
is responsible for the coupling between the monomers which results
from a Morse interaction potential of depth $D$, where $\alpha$ is the
range parameter and the parameter $l_0$ denotes the equilibrium
distance between the monomers. Throughout the paper we chose
$l_0=0.5$, i.e. the equilibrium distance amounts to half the
length of one period of the washboard potential.

The equations of motion are
\begin{eqnarray}
 \ddot{q}_1&=&-\sin(2 \pi q_1)+F(1-\tanh(q_1-q_0))\nonumber\\
&+&\alpha D (1-\exp[-\alpha(q_2-q_1-l_0)])\exp[-\alpha (q_2-q_1-l_0)]\label{eq:q1}\\
 \ddot{q}_2&=&-\sin(2 \pi q_2)+F(1-\tanh(q_2-q_0))\nonumber\\
&-&\alpha D (1-\exp[-\alpha(q_2-q_1-l_0)])\exp[-\alpha (q_2-q_1-l_0)]\label{eq:q2}\,.
\end{eqnarray}
The interaction strength between the two monomers is effectively
determined by the product $\alpha D$. For $\alpha D=0$ the system
decouples into two integrable subsystems and the dynamics is
characterised by individual regular monomer motions in the
washboard potential. For nonzero $\alpha D$ the dynamics is no
longer integrable. To prevent unphysical events in which the left
monomer overtakes the right one, a sufficiently strong coupling
between them is required. The  choice $\alpha=3$ and
$D\in[0.5,3]$ ensures that. On the other hand the effective
coupling strength, $\alpha D\in[1.5,9]$, is then too large by far
to treat the coupling using a perturbational approach. We therefore
resort to a numerical analysis of the coupled monomer dynamics.

Let us briefly discuss the phase space structure corresponding to
the dynamics in the tilted washboard potential. In the range
$-\infty <q \lesssim 10$ the tilt force, $-dU_1/dq$, is
effectively of strength $2F$. For uncoupled monomers ($\alpha
D=0$)  there exist saddles at ${q}_s^k=0.5+k-\arcsin
(2F)/(2\pi)$ and centers at ${q}_c^k=k-\arcsin (2F)/(2\pi)$
for integer values $k$.
For very small tilt strength $F\lesssim 0.01$ the barrier height
of the washboard potential, given by the difference between the
energy of the saddle and the center, is virtually equivalent to
those of the corresponding unbiased system with $F=0$, i.e.
$E_b\simeq 1/\pi$.

\section{Particle current}\label{section:current}

In this section we consider the emergence of a particle current.  The
initial positions of the monomers are taken as $-q_1(0)=q_2(0)=0.25$,
so that the dimer is contained in one of the wells of the washboard
potential and, for the weak tilt strength $F=0.01$ used
throughout the paper, is initially situated in virtually the
lowest energy dimer configuration compatible with the bond length
$l_0$. The dimer initially has potential energy
$E_{pot}=\sum_{n=1,2}[U_0(q_n)+U_1(q_n)]=0.3221$ which is of the order
of the barrier energy $E_b\simeq 1/\pi\simeq 0.3183$ of the washboard
potential. Note that since the bond between the monomers is initially
undistorted, the contribution from the Morse potential energy
$H_{int}$ to the system's initial potential energy is zero. The initial
kinetic energy of the dimer is taken as $E_{kin}=0.1234$. Crucially
the total energy $E_{total}=0.4455<2E_b$ is not sufficient that the
two monomers can escape simultaneously from a well of the washboard
potential. In order for directed motion of the dimer to occur
at all, cooperation between the monomers, in the form of
appropriately coordinated energy exchanges, is required.

Particle transport is assessed quantitatively by the mean momentum,
viz. the {\it current}, which is defined as the time average of the
ensemble averaged momentum, i.e.
\begin{equation}
p= \frac{1}{T_s}\,\int_0^{T_s} dt^{\prime} \langle{p}_{1}(t^{\prime})
+{p}_{2}(t^{\prime})\rangle \,,
\label{eq:current1}
\end{equation}
with simulation time $T_s$ and with the ensemble average given by
\begin{equation}
\langle {p}_{i}(t)\rangle=\frac{1}{N}\sum_{n=1}^N\,{p}_{i,n}(t)\,,\qquad i=1,2\,.\label{eq:current2}
\end{equation}
Here $N$ denotes the number of particles constituting the
ensemble. For the computation of the ensemble average, trajectories
belonging to $N=2\times10^5$ values of the pair of initial momenta
$(p_1(0),p_2(0))$ are taken. These initial values are uniformly
distributed on an iso-energetic ring in the $p_1-p_2-$plane such that
the relation
\begin{equation}
2E_{kin}=p_1^2+p_2^2\label{eq:ring}
\end{equation}
is fulfilled. 
Notice the symmetry $p_i\leftrightarrow -p_i$ and $i=1,2$. Hence there is no bias contained in the ensemble of initial conditions.
The simulation time interval is $T_s=10^5$ being
equivalent to almost $4\times10^4$ the period duration for
harmonic oscillations near the bottom of a potential well.

In what follows we vary the depth of the Morse potential, $D$,
playing, for fixed $\alpha=3$, the role of the coupling parameter.
The dependence of the current, defined in Eqs.~(\ref{eq:current1}) and
(\ref{eq:current2}), on the value of $D$ is shown in
Fig.~\ref{fig:current}. 
For values $D\lesssim 3.9$ the current exhibits variations and even vanishes for $D=1.1$. Interestingly the current effectively grows for $d\gtrsim 1.2$. Finally the current rises rapidly and monotonically in the range
$3.9\lesssim D \lesssim 4.5$ and effectively saturates at a high level for
$D\gtrsim 4.5$.
\begin{figure}
\includegraphics[scale=0.4]{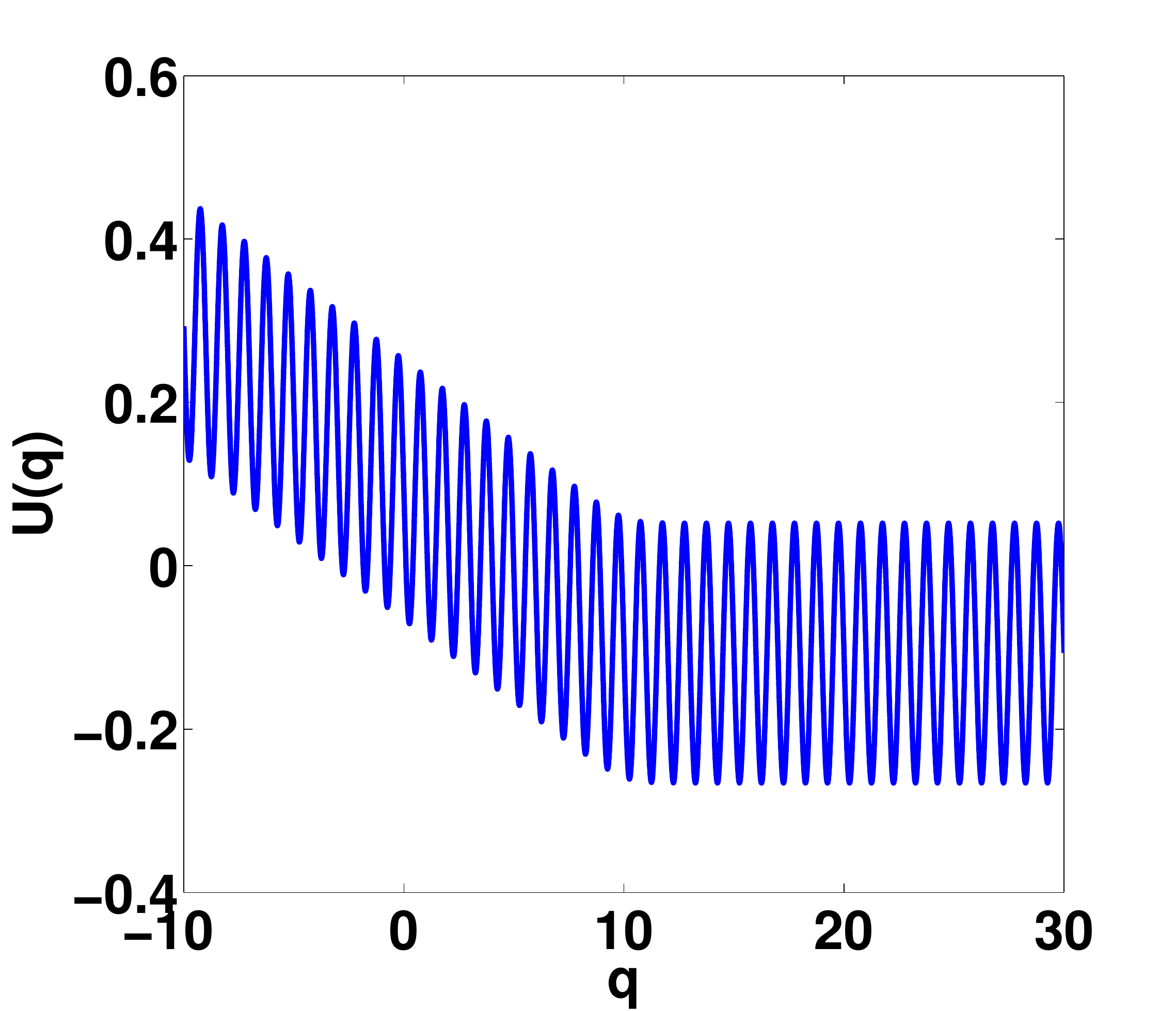}
\caption{ The potential $U(q)=U_0(q)+U_1(q)$ for parameter values $F=0.01$ and $q_0=10$.} \label{fig:pot}
\end{figure}
We emphasise that the amplitude of the tilt force $F=0.01$ is too
small to alter the washboard potential significantly compared to the
case without tilt.  In fact the influence of the tilt force is
sufficiently small that, for example, the potential barrier
immediately to the right of the initial well is lowered by a
mere $7.5\%$ and hence the induced bias is very weak. 

In the following we illustrate the complex solution behaviour of
system (\ref{eq:q1}),(\ref{eq:q2}) and the implications for the contribution to the net current.
Fig.~\ref{fig:time} shows the temporal
evolution of the coordinates $q_1(t)$ and $q_2(t)$ for four
different values of $D$ but for the same initial condition leading to 
various types of solutions.
\begin{figure}
\includegraphics[scale=0.4]{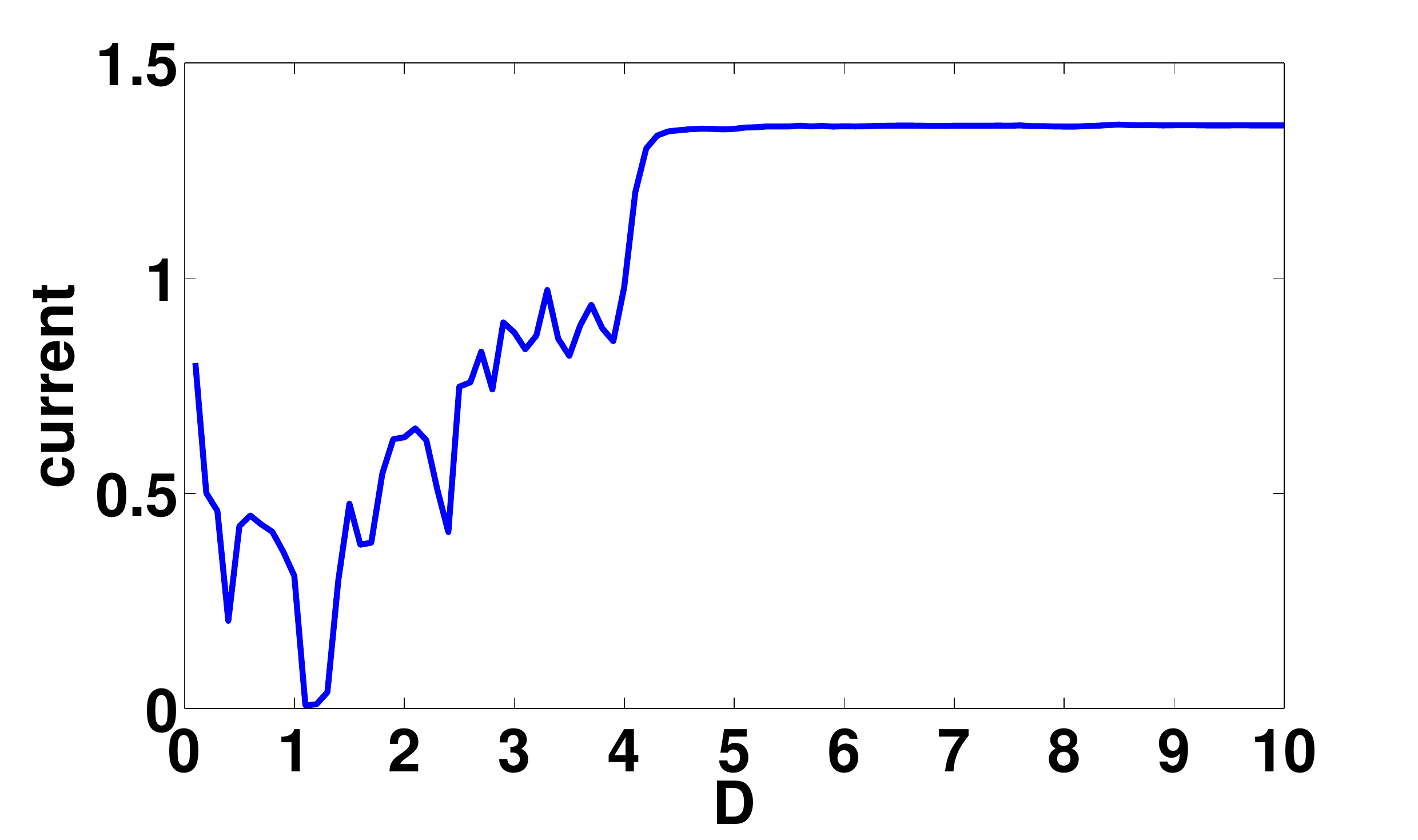}
\caption{Current as a function of $D$ 
in dimensionless units.
The remaining parameter values are given by: $\alpha=3$, $F=0.01$ and $q_0=10$.} \label{fig:current}
\end{figure}
For the low value $D=0.5$ the coordinates perform small-amplitude
oscillations around their respective starting value (see the upper
left panel of Fig.~\ref{fig:time}). Thus the monomers remain trapped
in the potential well and the contribution to the net current is
zero. In contrast for $D=1$ we observe that after a finite period of
chaotic but bounded dimer motion the bond between them breaks. As a
result the right monomer (with index $n=2$) is released and due to the
(still acting) tilt force accelerated into the region of higher
momenta while its left counterpart (with index $n=1$) becomes again
trapped in a potential well. The subsequent regular dynamics is
characterised by different motions of the monomers, namely that of the
right monomer moving rightwards (rotations) in the asymptotic region
and the left monomer performing bounded oscillations in a potential
well (librations). In this case, the directed motion of the right
monomer gives a contribution to the net current. We stress that, after
such fragmentation, reformation of a bound state dimer from the two
isolated monomers is excluded. Notice that the possibility of bond
breaking allows for transient chaos
\cite{Tel},\cite{Tel1},\cite{Zaslavsky}.
\begin{figure}
\includegraphics[scale=0.4]{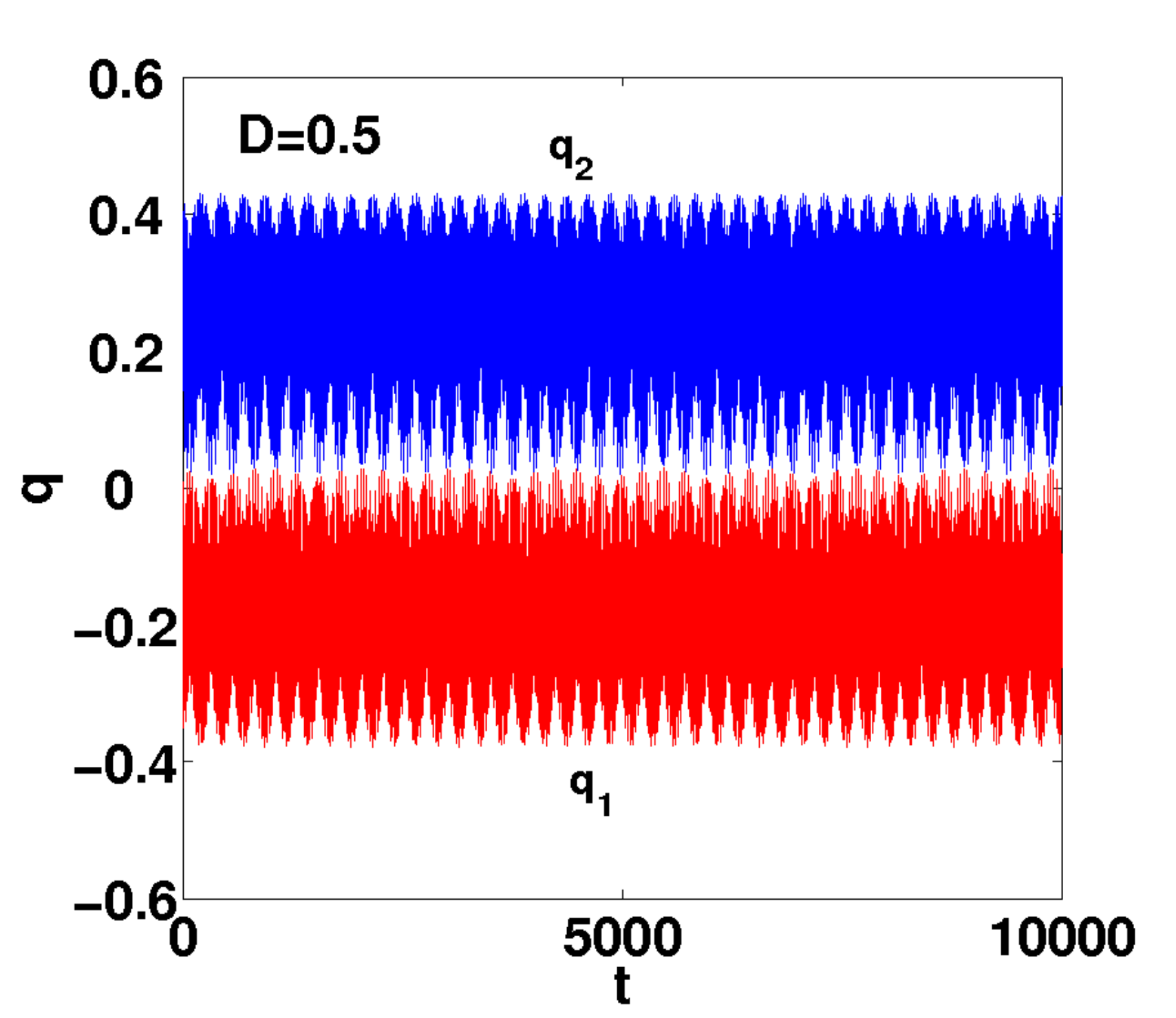}
\includegraphics[scale=0.4]{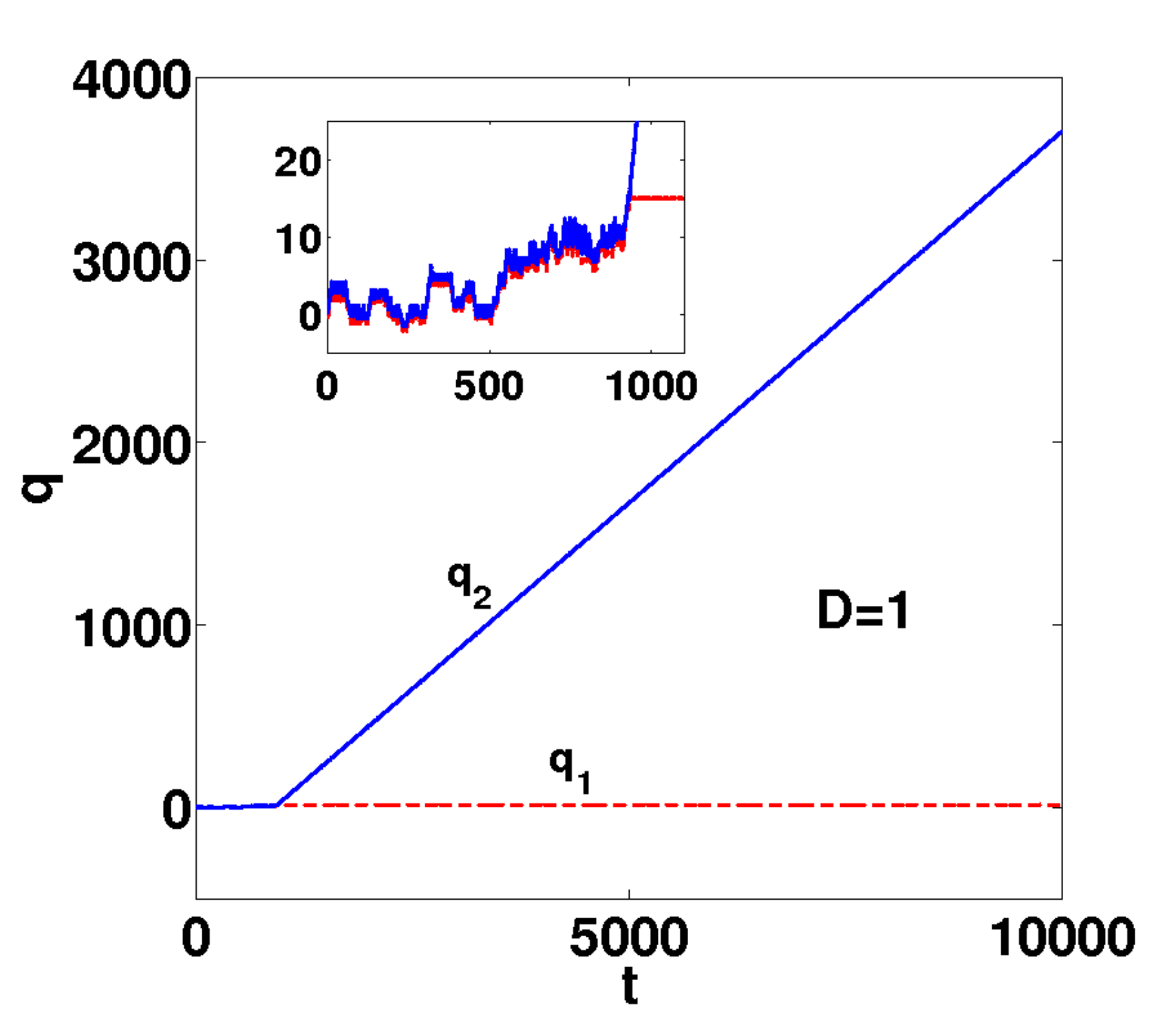}
\includegraphics[scale=0.4]{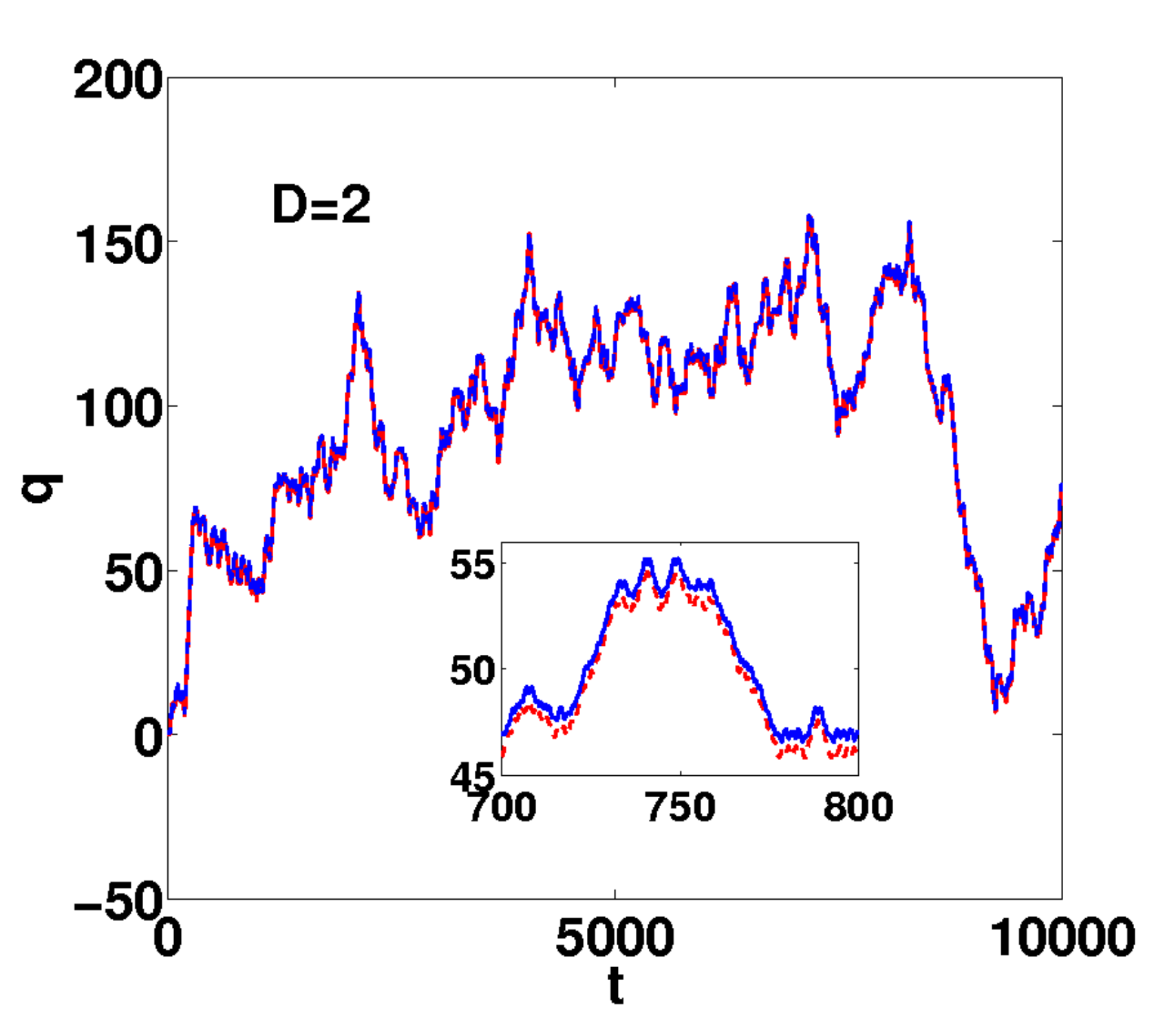}
\includegraphics[scale=0.4]{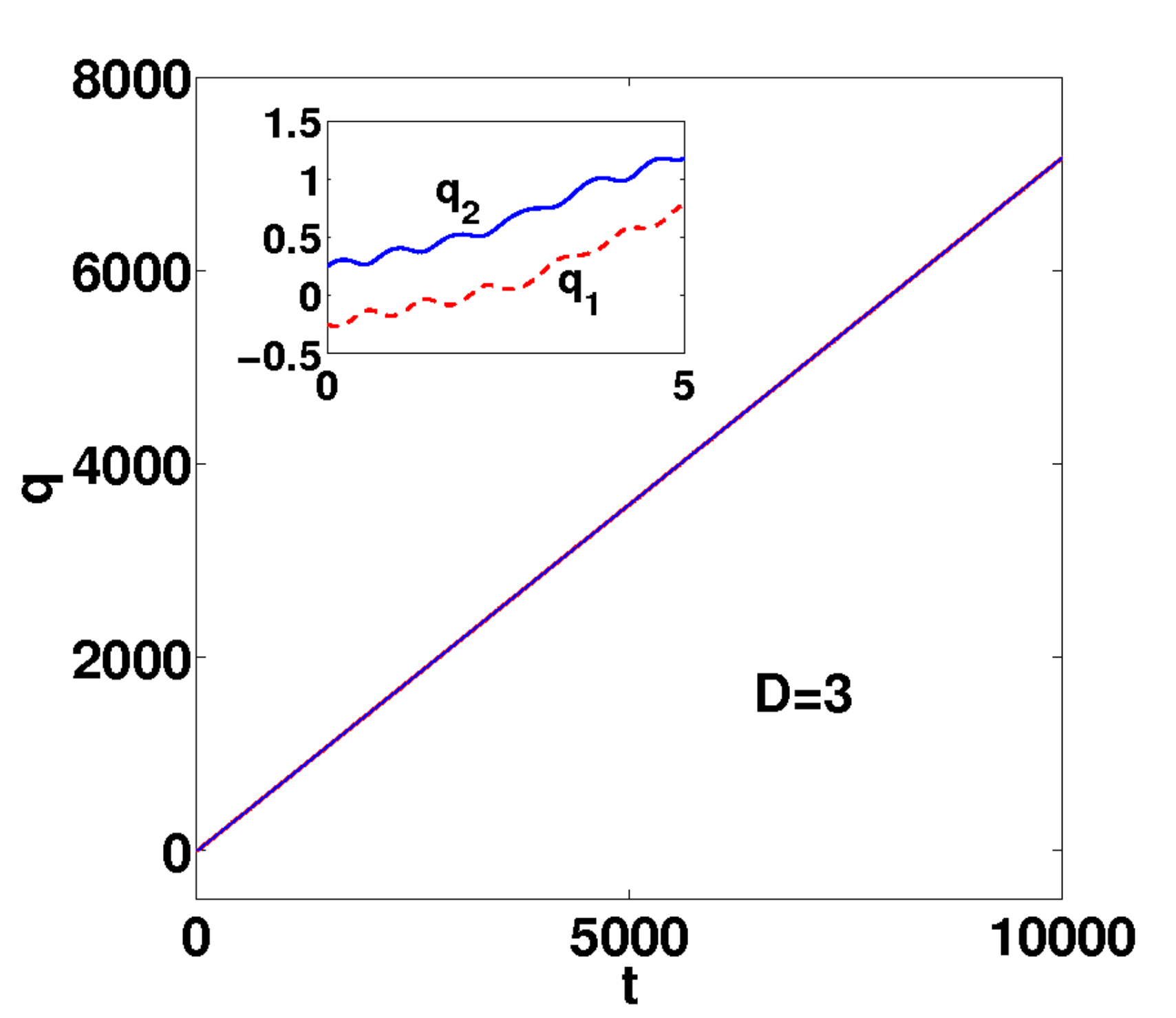}
\caption{ Four qualitatively different scenarios, illustrated by the
time evolution of individual coordinates with initial
conditions $p_1(0)=-0.2100$ and $p_2(0)=0.4502$ for the values
of $D$ indicated in the plots. The remaining parameter values
are given by: $\alpha=3$, $F=0.01$ and $q_0=10$.
The insets show details of the evolution.} \label{fig:time}
\end{figure}

Increasing the coupling parameter further to $D=2$ the
chaotic coupled monomer dynamics involves irregular phases where
the motion changes from forward to backward and vice versa 
in the manner of L{\'e}vy flights 
\cite{Levy}
in the whole simulation interval. Nevertheless the net motion proceeds
to the right. Notice that the bond remains unbroken, and hence the
dimer intact, in this case.

Interestingly for a high value $D=3$ the dimer manages quickly to
escape from the potential well. (We remark that for some other initial
conditions we observed first a longer transient of still bounded but
chaotic motion before the escape eventually took place.) Furthermore,
as the inset reveals, the two monomers perform out-of-phase motion,
viz. the length of the bond between them alternately (slightly)
decreases and increases. This is associated with such well-coordinated
energy exchanges between the monomers that firstly the right-hand
monomer overcomes the potential barrier and reaches the adjacent well
on the right, subsequently the left monomer follows, and so on. We
underline that prior to its arrival in the asymptotic region the
trajectory passes through a chaotic transient to adopt regular dynamics in
the asymptotic region.  It is the asymptotic
vanishing of the tilt force that makes transient chaos possible.
Clearly the directed dimer motion contributes with significant weight
to the net current. In particular, the dimer moves with higher velocity
for $D=3$ than the escaped monomer does for $D=1$.

To summarise briefly: we distinguish between four qualitatively
different transport scenarios:\\ \noindent (i) The dimer remains
trapped inside the starting potential well and hence, there results no
contribution to the net current.\\ \noindent (ii) The dimer escapes
from the starting potential and undergoes subsequently diffusive
motion with no substantial contribution to the net
current.\\ \noindent (iii) Directed energy transfer from the left
monomer to the right one leads not only to fragmentation but also to
such a high energy gain of the right monomer that it can undergo
directed motion to the right. This {\it individual} directed motion
yields a considerable contribution to the net current.\\ \noindent
(iv) Appropriately coordinated energy redistribution between the
monomers leads to directed {\it collective} motion such that first the
right monomer performs a transition from one potential well into the
next one to the right and afterwards the left monomer follows. Notice
that this corresponds to repeated detrapping-trapping
transitions. This scenario is optimal in the sense that it enables
both monomers to perform consecutive step-wise escapes from the
potential wells. Since the total amount of energy does not suffice for
a simultaneous escape of both monomers they must necessarily share
their energy cooperatively in order to achieve escape at all. This
almost-periodic energy exchange between the monomers corresponds in
phase space to motion near a stable period-one fixed point (see
further in \ref{section:potential}).  Notably the directed chaotic
motion persists even in the asymptotic region where there is no bias
anymore. Furthermore, the resulting velocity is higher than in case
(iii) and so is the contribution to the net current.  We mention that,
apart from the ideal situation of ongoing directed motion, for other
initial conditions the dimer performs directed long-distance motion in
a restricted time interval at the end of which non-directed diffusive
motion as in case (ii) follows.

We emphasise that the scenarios shown in Fig.~\ref{fig:time} are not
necessarily representative of the dynamics of all initial conditions
at the respective values of the coupling strength $D$; at $D=3$, for
example, scenarios (iii) and (iv) both occur (for different initial
conditions).  Concerning the current we therefore remark that, at each
fixed value of the coupling strength $D$, each of the transport
scenarios (i)-(iv) will, if present, contribute with different weight
to the ensemble average for the current, which results in the complex
behaviour seen in Fig.~\ref{fig:current}.

\section{Phase space dynamics}\label{section:space}

\begin{figure}
\includegraphics[height=6cm, width=8cm]{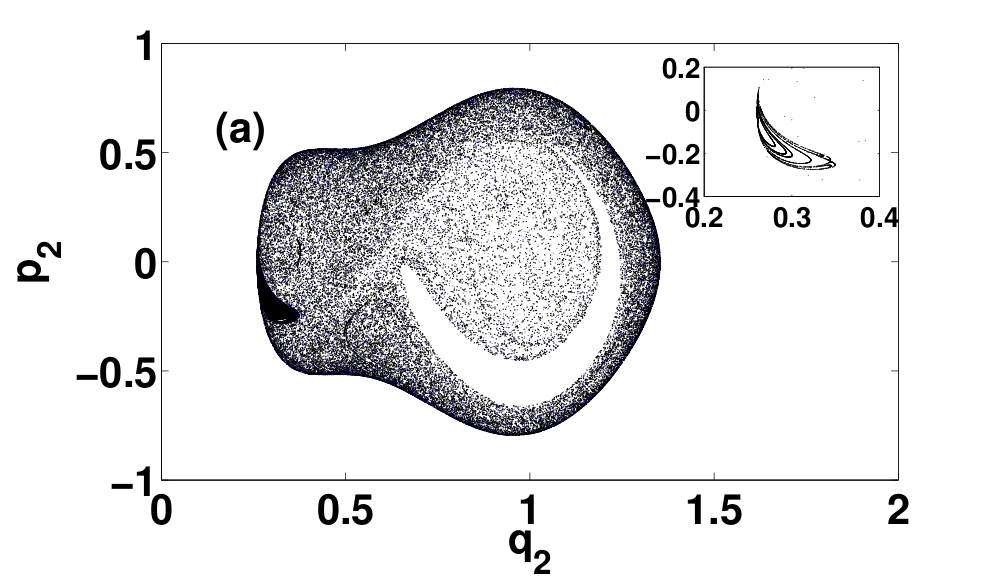}
\includegraphics[height=6cm, width=8cm]{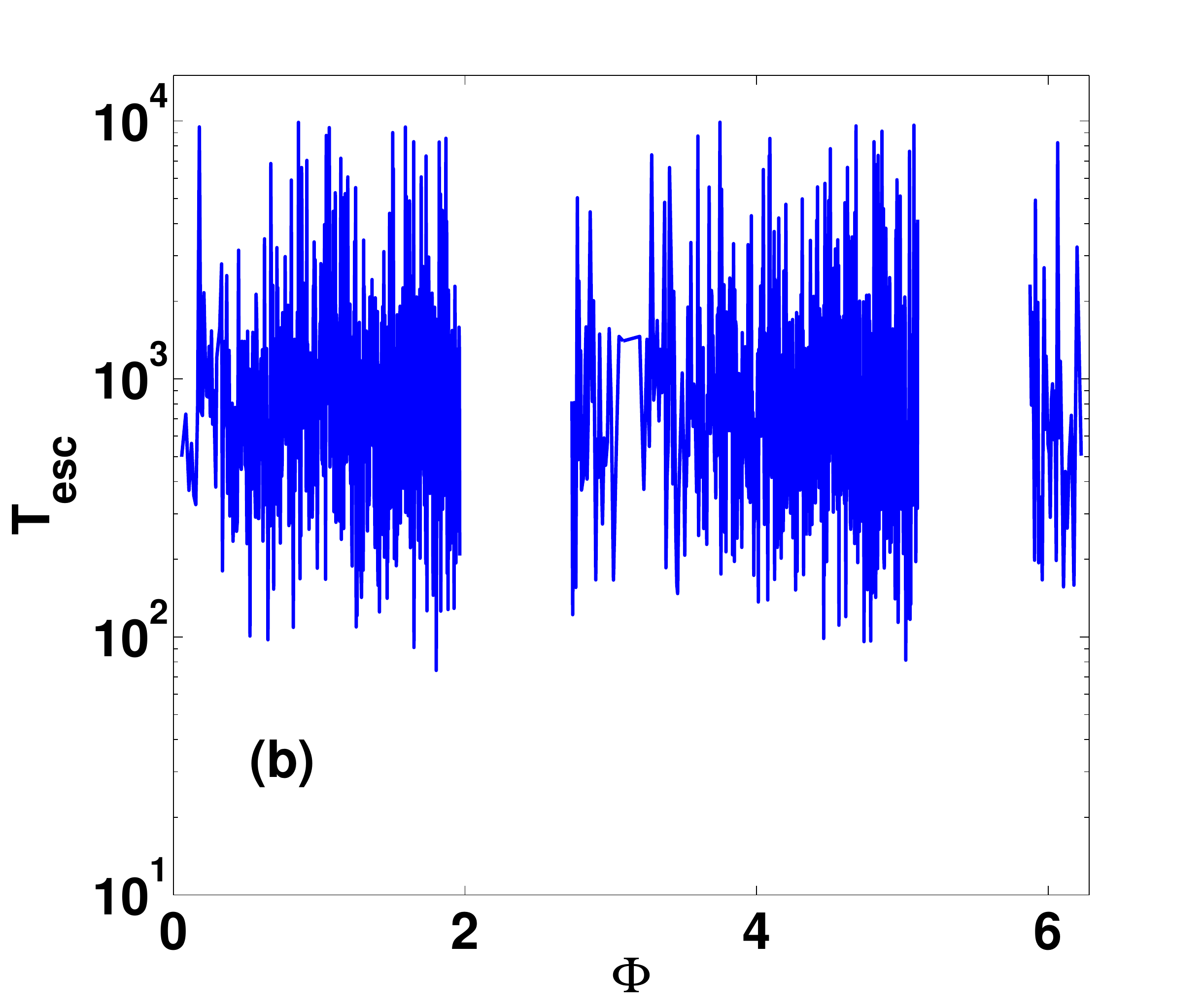}
\includegraphics[height=6cm, width=8cm]{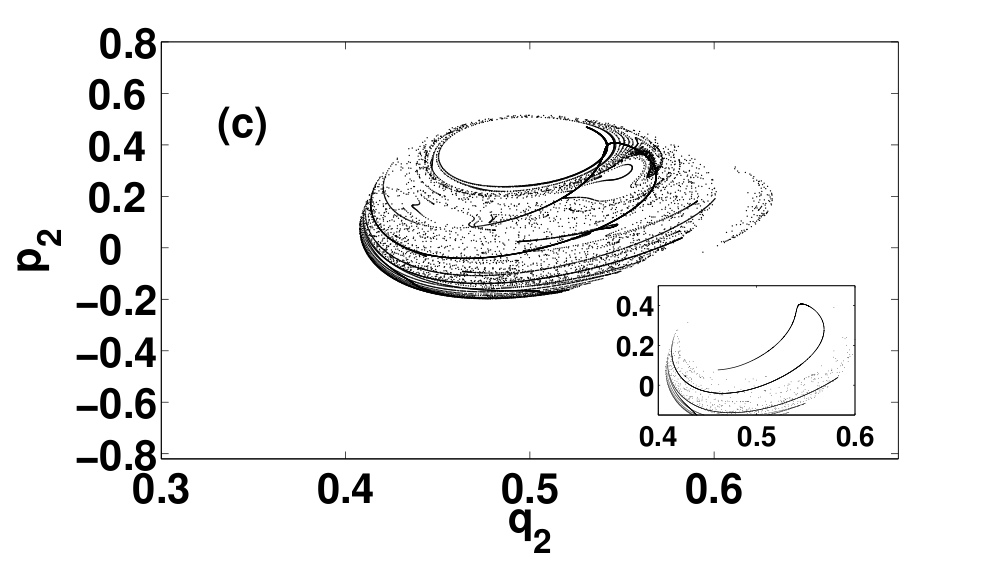}
\includegraphics[height=6cm, width=8cm]{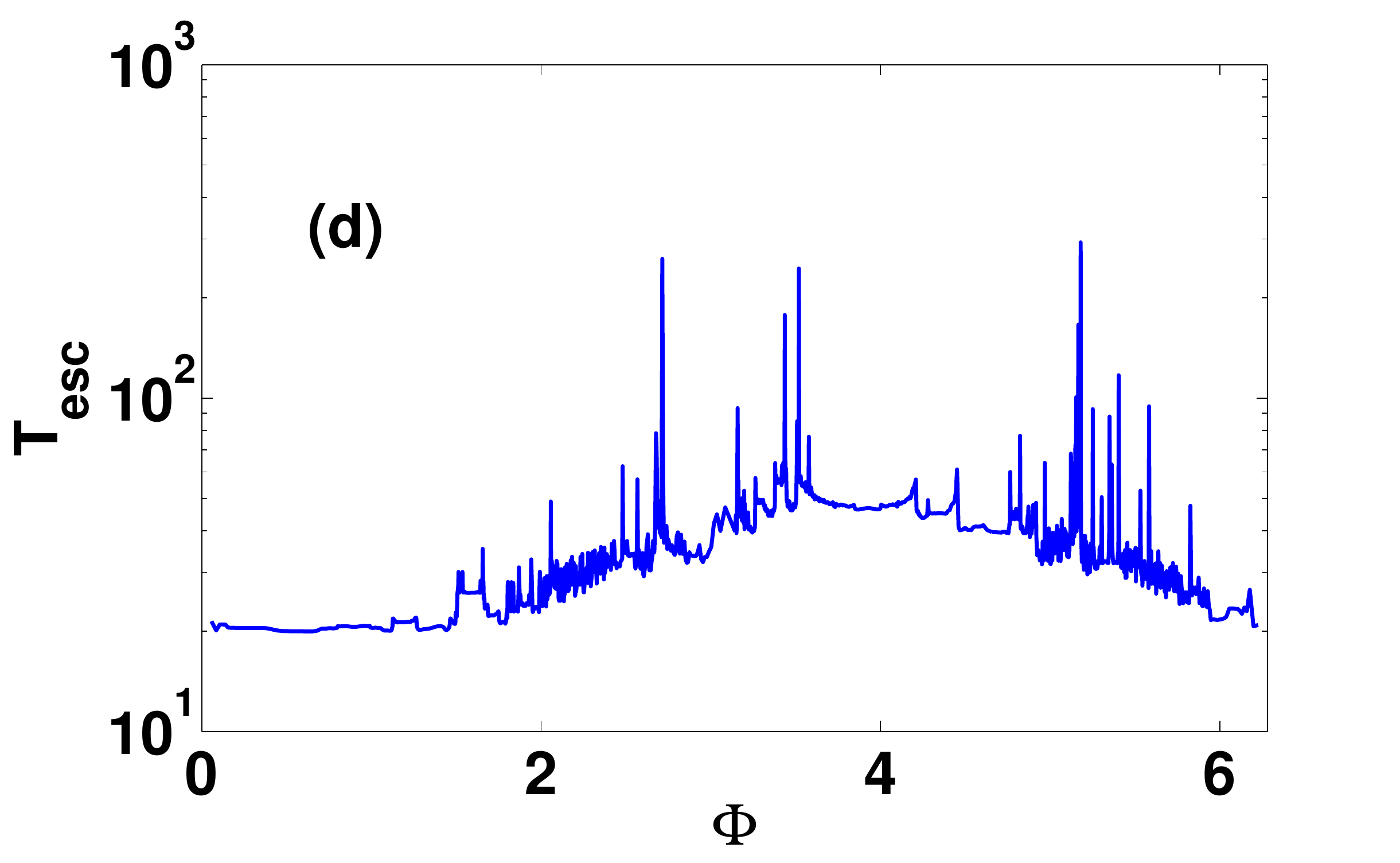}
\caption{ PSS represented in the $p_2-q_2$ plane (a) and (c) and
  escape times versus the angle $\Phi=\tan^{-1}(p_2(0)/p_1(0))$ shown
  in (b) and (d).  Upper (resp. lower) row: $D=0.5$ (resp. $D=3$).
  The remaining parameter values are given by: $\alpha=3$, $F=0.01$
  and $q_0=10$. An ensemble of $10^4$ initial conditions fulfilling
  relation (\ref{eq:ring}) with $E_{kin}=0.1234$ is
  used.} \label{fig:escape}
\end{figure}

In order to discuss the corresponding dynamics taking place on the
three-dimensional energy hypersurface in the four-dimensional
phase space we introduce the following Poincar\'{e} surface of section
(PSS)
\begin{equation}
\Sigma=\left\{\,p_2, q_2|q_1=0, p_1>0\,\right\}\,.\label{eq:PSS}
\end{equation}
In Fig.\,$4$\,(a) and (c) we depict the PSS for $D=0.5$ and $D=3$, respectively using 
an ensemble of $10^4$ initial conditions fulfilling the relation (\ref{eq:ring}) with $E_{kin}=0.1234$.
The corresponding right panel presents
the {\it escape time function}, defined as the time it takes the
right monomer to reach the position $q_2=10$,  as a function of
the angle
\begin{equation}
\Phi=\tan^{-1}(p_2(0)/p_1(0)).
\end{equation}
For $D=0.5$ there are
two wide regions
on the $\Phi-$axis for which no escape happens at all. 
This is the case when the monomers start with initial momenta of
approximately equal magnitude but of opposite sign, i.e. around
$\Phi\simeq 3\pi/4$ and $\Phi \simeq 7\pi/4$.  In both cases the
resulting regular motion is associated with the stable island centered
at $(p_2,q_2)\simeq(-0.2,0.4)$ in the corresponding PSS (displayed in
the inset of Fig.~\ref{fig:escape}\,(a)). The physical reason for the
appearance of regular trapped motion in these cases is the fact that
the initial velocity of the center of mass of the dimer is virtually
zero --- a situation that remains virtually unchanged due to the
symmetry of the washboard potential, the weakness of the tilt, and the
small interaction term.
An examination of the escape time function at various scales
reveals that, except for the two windows of no-escape, the escape time
depends sensitively on changes of the initial momenta. The
trajectories attributed to escaping monomers are contained in the
extended chaotic sea which densely fills the majority of the
energetically-accessible regions on the PSS (except for an infinite
set of smaller islands of stability not recognisable on the scale of
the PSS).  A crescent-moon-shaped region within the chaotic sea
remains empty on the PSS because it is not energetically accessible.
In more detail, there exist chaotic invariant sets consisting of
homoclinic and heteroclinic tangles which induce a fractal set of
singularities into the escape time function
\cite{Ott}-\cite{Barr}. The singularities arise at those points where
the stable manifolds of unstable periodic orbits intersect the set of
initial data with the effect that the corresponding trajectories
become trapped for arbitrarily long times in a chaotic invariant
set. It is therefore impossible to fully resolve the behaviour of the
escape time function whose singularities form a fractal set with
measure zero.

Interestingly, for increased coupling strength $D=3$ the interaction
between the monomers is strong enough that fully developed chaos
results and the windows of no-escape obtained in the previous case of
$D=0.5$ vanish.  In comparison, the escape times are mostly shorter by
far for $D=3$ than for the preceding case $D=0.5$.  The associated PSS
elucidates these differences in the escape process.  (We remark that
according to the condition in (\ref{eq:PSS}) only those trajectories
for which the left monomer is still in the starting potential well
contribute to the PSS.)  Comparing the cases $D=0.5$ and $D=3$ one
infers that in the former case the chaotic sea engulfs far more area
(extending along the $q_2-$axis over the range of the starting
potential well together with its neighbour to the right) despite the
existence of the (small) stable island (inset in
Fig.~\ref{fig:escape}(a)). Moreover the fact that the PSS is more
densely populated for $D=0.5$ than in the case $D=3$ indicates that
trajectories spend longer times in the potential well(s) for $D=0.5$
before they manage to escape. Furthermore, for $D=3$ some trajectories
follow directly the unstable manifold associated with a chaotic saddle
appearing as a winding curve emanating from the region around $p_2=0,
q_2\simeq 0.5$ (inset in Fig.~\ref{fig:escape}\,(c)). Further details
are given in Section \ref{section:potential}. This provides a
mechanism for fast escape into the range of large coordinates which
happens particularly for initial values lying in the range
$0<\Phi\lesssim 1.5$.  Nonetheless there remains a large portion of
trajectories that dwell for some time in the starting region before
they escape in the direction of the asymptotic region. The dwell time
depends sensitively on the initial conditions. On the other hand,
escape does not necessarily imply sustained directed transport.

\begin{figure}
\includegraphics[height=6cm, width=8cm]{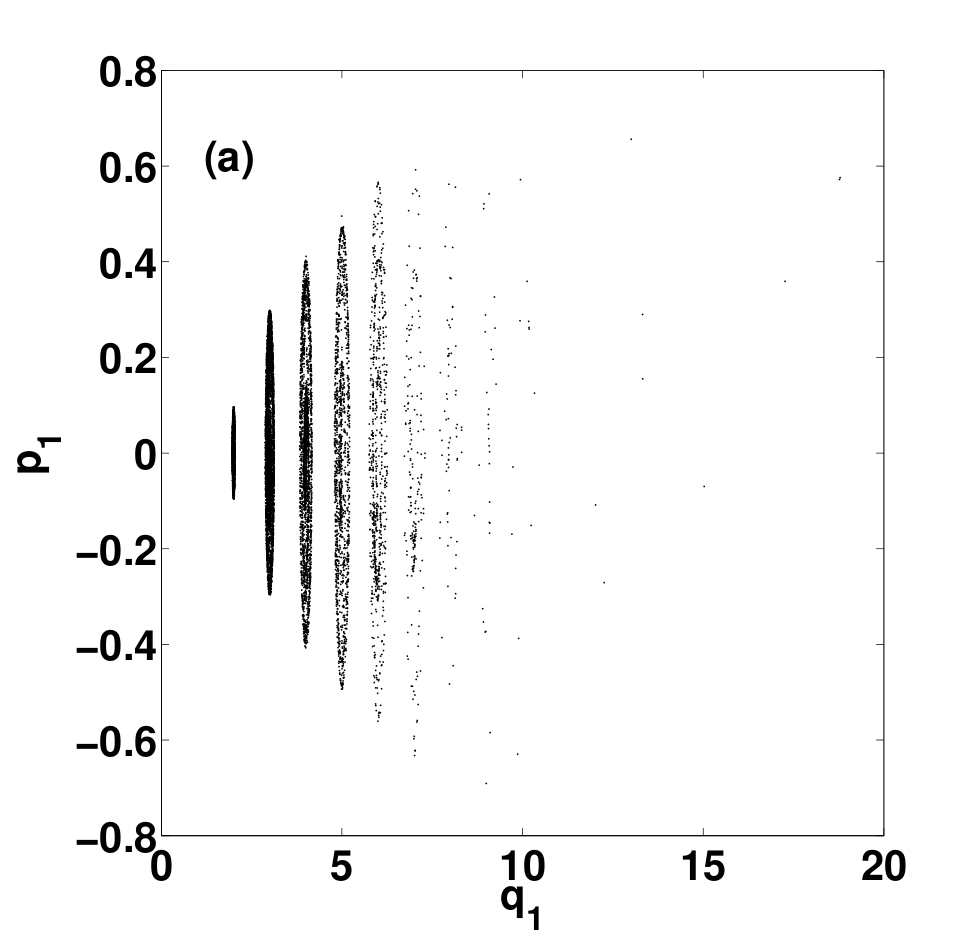}
\includegraphics[height=6cm, width=8cm]{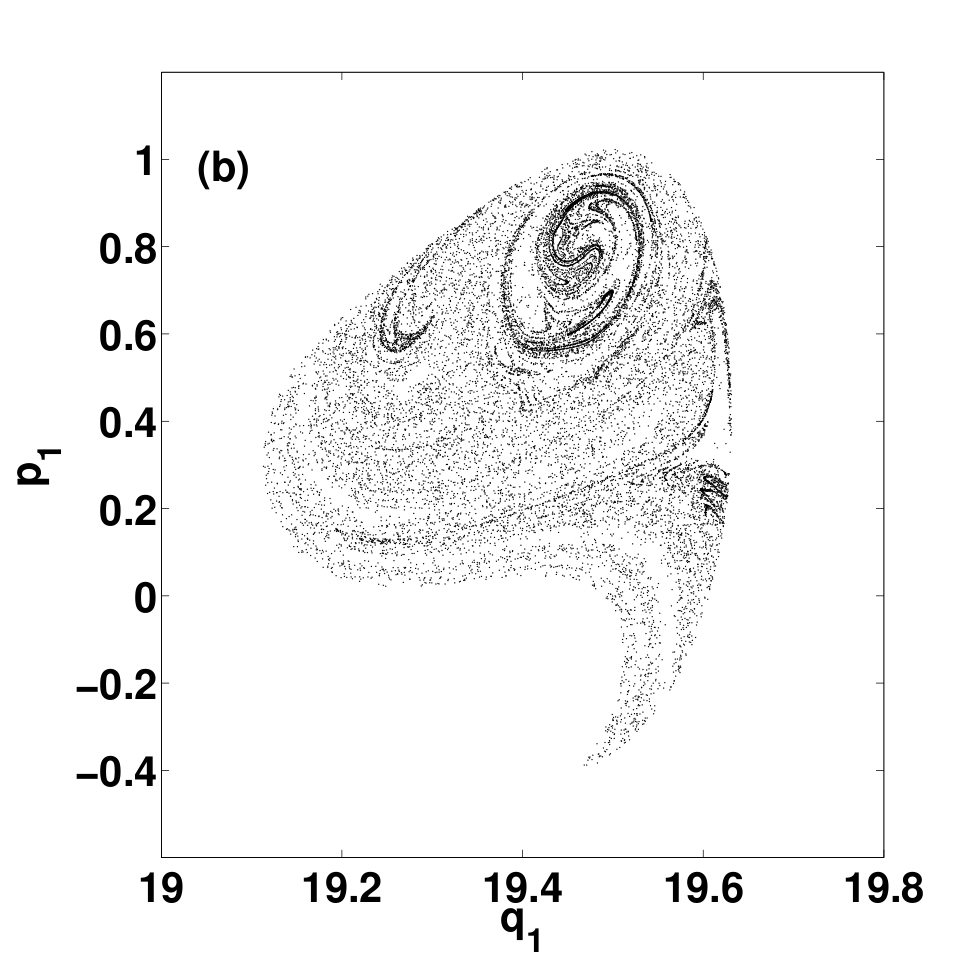}
\caption{Sections taken when the coordinate of the right monomer reaches
  $q_2=20$ with $10^4$ initial conditions from an ensemble satisfying
  the relation (\ref{eq:ring}) with $E_{kin}=0.1234$.  (a) (resp. (b)):
  $D=0.5$ (resp. $D=3$). The remaining parameter values are given by:
  $\alpha=3$, $F=0.01$ and $q_0=10$.}
\label{fig:psd=0.520}
\end{figure}

For further characterisation of the escape dynamics PSS are presented
in the $p_1-q_1-$plane using the intersection condition $q_2=20,
p_2>0$ yielding a snap shot of the ensemble dynamics in the asymptotic
region.  Note that at such a large distance from the starting point
located at $q_2(0)=0.25$, the right monomer experiences an unbiased
potential. The ensemble of initial conditions is the same as the one
used for Fig.~\ref{fig:escape}.  The PSS for $D=0.5$ is depicted in
Fig.~\ref{fig:psd=0.520}\,(a) and shows that advancing right monomers
leave the overwhelming majority of their left counterparts behind
distributed over various potential valleys where they perform trapped
motion.  Clearly for the fairly low interaction potential depth
$D=0.5$ the bond between the monomers easily breaks. In contrast, for
the comparatively large interaction potential depth, $D=3$, the bond
between the monomers remains intact for the entire time, as seen in
the right panel of Fig.~\ref{fig:psd=0.520}. Hence, left monomers
travel the full distance in unison with their right counterparts.

It is illustrative to consider the distribution of the momenta of the
right monomers at the moment when they reach the position
$q_2=20$. For $D=0.5$ the momenta are narrowly distributed around the
peak value $p_2\simeq 0.92$ (not shown).  In this case the particle
transport is dominated by directed motion of right monomers after
fragmentation (cf. scenario (iii) above). For $D=3$ there results a
broad momentum distribution in an interval matching that covered by
the $p_1-$values in the right panel of Fig.~\ref{fig:psd=0.520}. This
indicates that the dynamics in the asymptotic region involves not only
directed motion but also itinerant motion as described above by the
diffusive-like scenario (ii).  Nevertheless the distribution of the
$p_2$ values attains a maximum at $p_2\simeq 0.81$, viz. for momenta
for which directed motion to the right proceeds.

\section{Motion of a particle in an effective two-dimensional potential}\label{section:potential}

\begin{figure}
\begin{tabular}{cc}
\includegraphics[width=6cm]{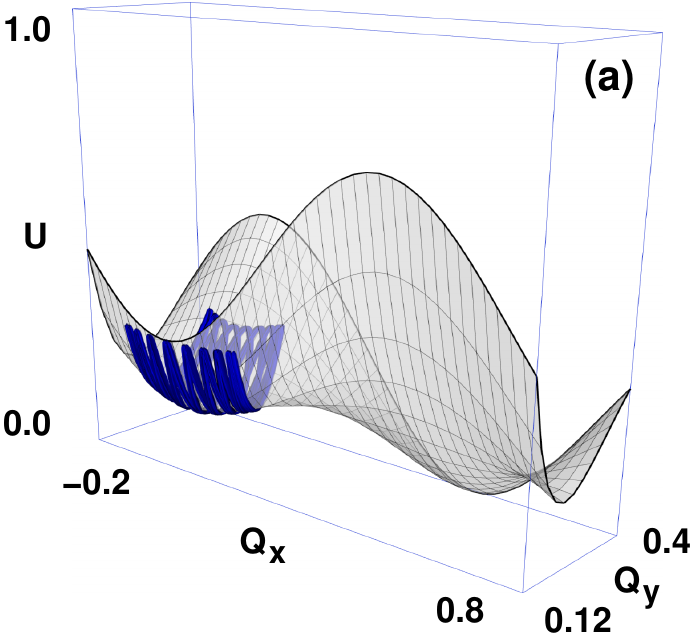}~&
\includegraphics[width=6cm]{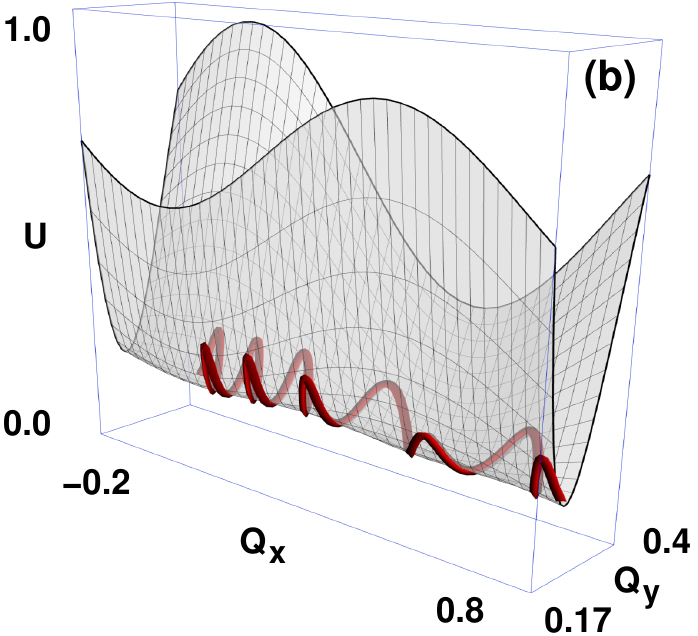}\\
\includegraphics[width=6cm]{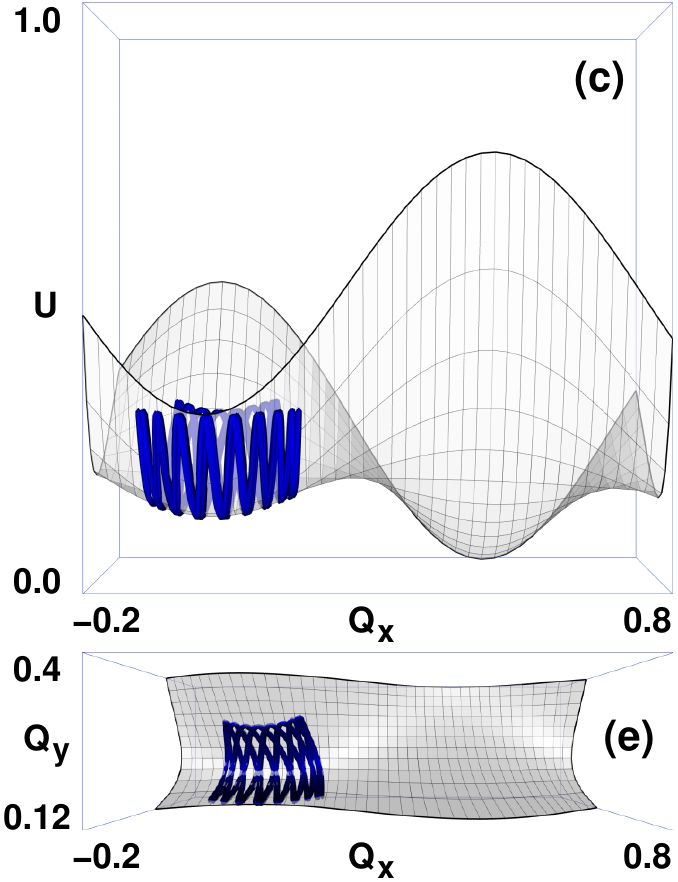}~&
\includegraphics[width=6cm]{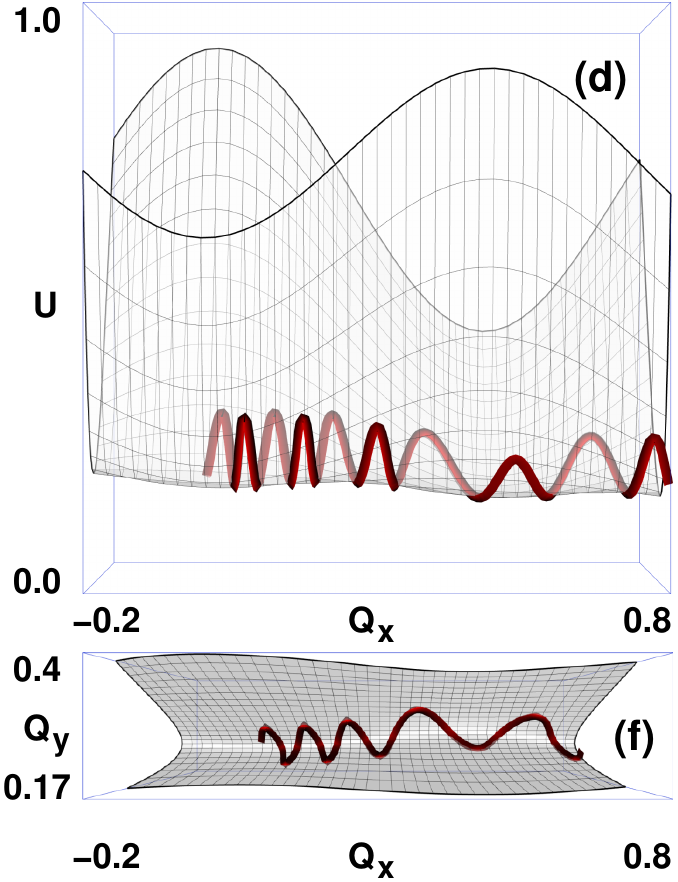}
\end{tabular}
\caption{(Colour online) Evolution of the trajectory in the
  two-dimensional potential energy landscape $U(Q_x,Q_y)$. The
  parameter values are $l_0=0.5$, $\alpha=3$. Left panels (a,c,e):
  trapped particle for $D=0.5$ and right panels (b,d,f): moving
  particle for $D=3$. For clarity, the middle (c,d) and bottom (e,f)
  rows show profile views ($Q_x,U$) and plan views ($Q_x,Q_y$),
  respectively. The steepness of the potential surface in the case
  $D=3$ (b,d,f) necessitates plotting the potential surface for a
  slightly smaller $Q_y$ range. The left and right panels are shown to
  the same scale in each case.} \label{fig:deltaU}
\end{figure}

\begin{figure}
\includegraphics[scale=0.5]{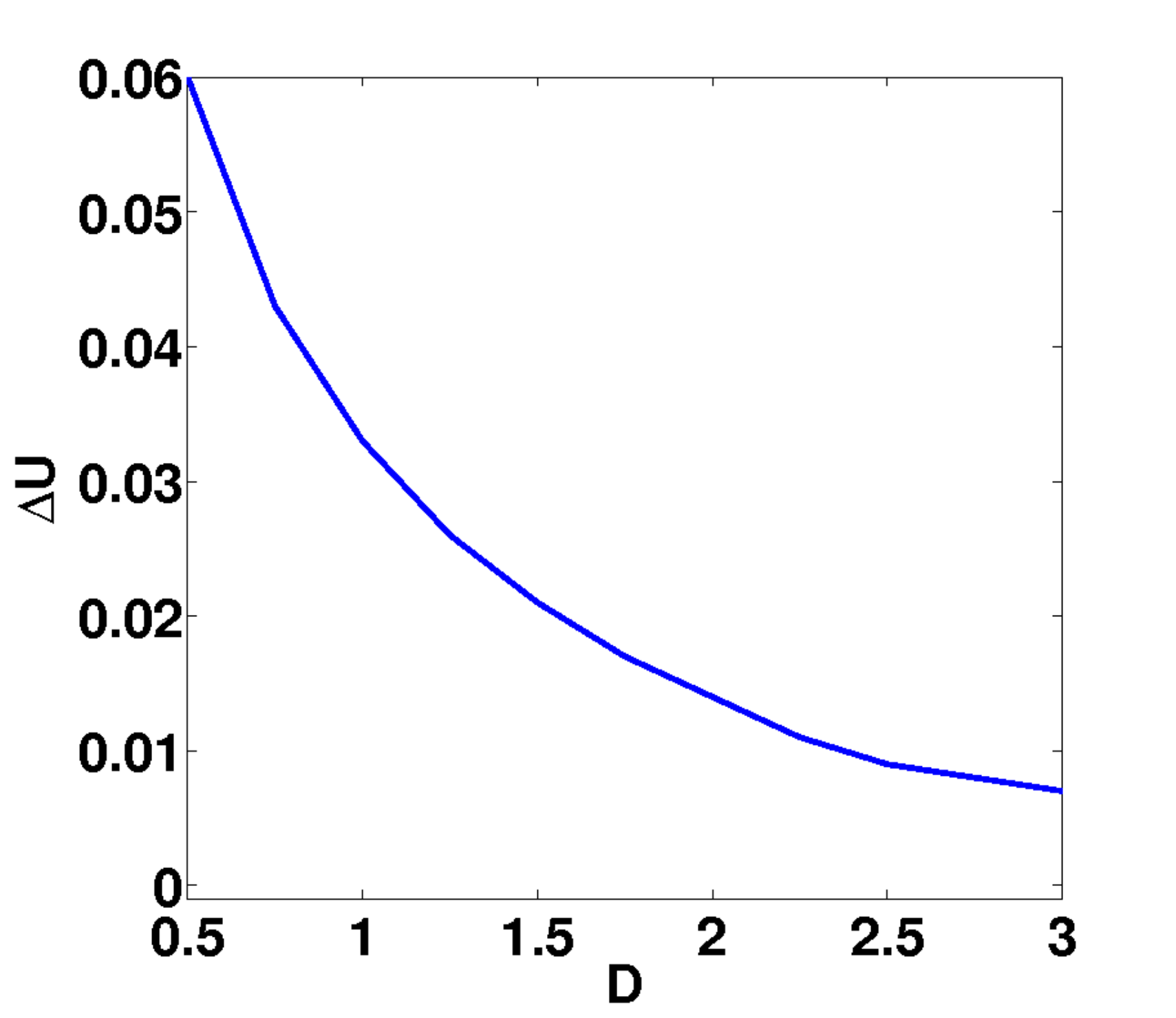}
\caption{(Colour online) Difference between the energy of the
potential minimum and the nearby saddle as a function of $D$. The
remaining parameter values are $l_0=0.5$, $q_0=10$, $F=0.01$, and $\alpha=3$.}
\label{fig:Fig2}
\end{figure}

To gain further insight into the coupled monomer dynamics it is useful
to perform the following canonical change of variables induced by the
generating function:
$S=\frac{1}{2}(q_1+q_2)P_x+\frac{1}{2}(q_2-q_1)P_y$ relating the old
and new variables as follows
\begin{eqnarray}
p_1&=&\frac{1}{2}(P_x+P_y)\,,\qquad p_2=\frac{1}{2}(P_x-P_y)\,,\\
Q_x&=&\frac{1}{2}(q_1+q_2)\,,\qquad Q_y=\frac{1}{2}(q_2-q_1)\,.
\end{eqnarray}
The coordinate $Q_x$ determines the position of the center of
mass (CM) of the dimer, accounting for translational
motion. Vibrations (V) of the dimer are described by $Q_y$.  The
Hamiltonian expressed in the new variables becomes
\begin{eqnarray}
H&=&\frac{1}{4}(P_x^2+P_y^2)-\frac{1}{\pi}\cos(2\pi Q_x)\cos(2\pi
Q_y)\nonumber\\ &+&\frac{D}{2} \left(1-\exp[-\alpha(2Q_y-l_0)]\right)^2\nonumber\\
&-&F(2Q_x-\log[\cosh(Q_x-Q_y-q_0)]-\log[\cosh(Q_x+Q_y-q_0)])\\
&\equiv&\frac{1}{4}(P_x^2+P_y^2)+U(Q_x,Q_y)
\,.\label{equation:Hnew}
\end{eqnarray}
The corresponding equations of motion, describing the effective motion of a
particle in a two-dimensional potential landscape $U(Q_x,Q_y)$, are given by
\begin{eqnarray}
\ddot{Q}_x&=&-2\sin(2\pi Q_x)\cos(2\pi Q_y)\nonumber\\
&+&F[2-\tanh(Q_x-Q_y-q_0)-\tanh(Q_x+Q_y-q_0)]\label{eq:qx}\nonumber\\
\ddot{Q}_y&=&-2\cos(2\pi
Q_x)\sin(2\pi Q_y)\nonumber\\
&-&2\alpha D(1-\exp[-\alpha(2Q_y-l_0)])\exp[-\alpha(2Q_y-l_0)]\nonumber\\
&+&F[\tanh(Q_x-Q_y-q_0)-\tanh(Q_x+Q_y-q_0)]\,.\label{eq:qy}
\end{eqnarray}
For $Q_x,Q_y\ll q_0$ the impact of the external tilt force matters
only in the first equation whereas the  Morse coupling enters only
in the second equation. The interaction between the $Q_x$ (CM) and
 $Q_y$ (V) degree of freedom (d.o.f.) results from parametric
modulations of the respective washboard potential force term. The
effective potential $U(Q_x,Q_y)$ is displayed in
Fig.~\ref{fig:deltaU} for the interaction potential
depths $D=0.5$ and $D=3$ respectively.
The superimposed trajectory, corresponding to the dynamics shown in
Fig.~\ref{fig:time}\,(a) and (c) respectively, starts close to the
potential minimum for $D=0.5$ situated at
$(Q_x,Q_y)=(0.014,0.173)$. The corresponding state of lowest energy is
denoted by $E_g$. There exists a nearby saddle, aforementioned in
Section \ref{section:space}, which for $D=0.5$ is located at
$(Q_x,Q_y)=(0.241,0.244)$ having energy $U_s$. In order to advance
towards higher $Q_x-$values in the two-dimensional potential landscape
the particle needs to overcome a potential barrier the height of which
is determined by $\Delta U=U_s-U_g$. Apparently for $D=0.5$ the
trajectory remains trapped inside the potential well
(cf. Fig.~\ref{fig:time}\,(a)).  This is mainly connected with the
relatively large-amplitude excursions in the $Q_y-$direction pointing
to rather pronounced bond stretching/compression. In fact, the major
part of the total energy is contained in the Morse interaction term,
viz. in the V-d.o.f., amounting to $70\%$. Thus there remains little
energy that can flow into the CM-d.o.f., hampering the translational
motion necessary to overcome the potential barrier. In contrast for
$D=3$, when the bond between the monomers is more rigid by far than
before, the Morse bond energy constitutes only a small amount of the
total energy. As a consequence the CM-d.o.f. possesses enough energy
that the trajectory easily overcomes the potential barrier and passes
from one well to a neighbouring one (right panel in
Fig.~\ref{fig:deltaU}). Moreover, along the line $Q_y=0.25$, that is
$-q_1=q_2=0.25$, there is no gradient of the potential in the
$Q_x-$direction (CM motion direction). Therefore a strong enough
interaction strength $D$ is advantageous for transport because it
confines the motion of the dimer along a narrow strip centered along
the line $Q_y=0.25$.  At the same time the height of the energy barrier
$\Delta U$ decreases with increasing interaction potential
depth as illustrated in Fig.~\ref{fig:Fig2}. Conclusively, motion of
the mean coordinate $Q_x$ from one potential well into the
neighbouring one is readily accomplished for large values of $D$ which
is reflected in a high current (see Fig.~\ref{fig:current} above).

\begin{figure}
\includegraphics[scale=0.3]{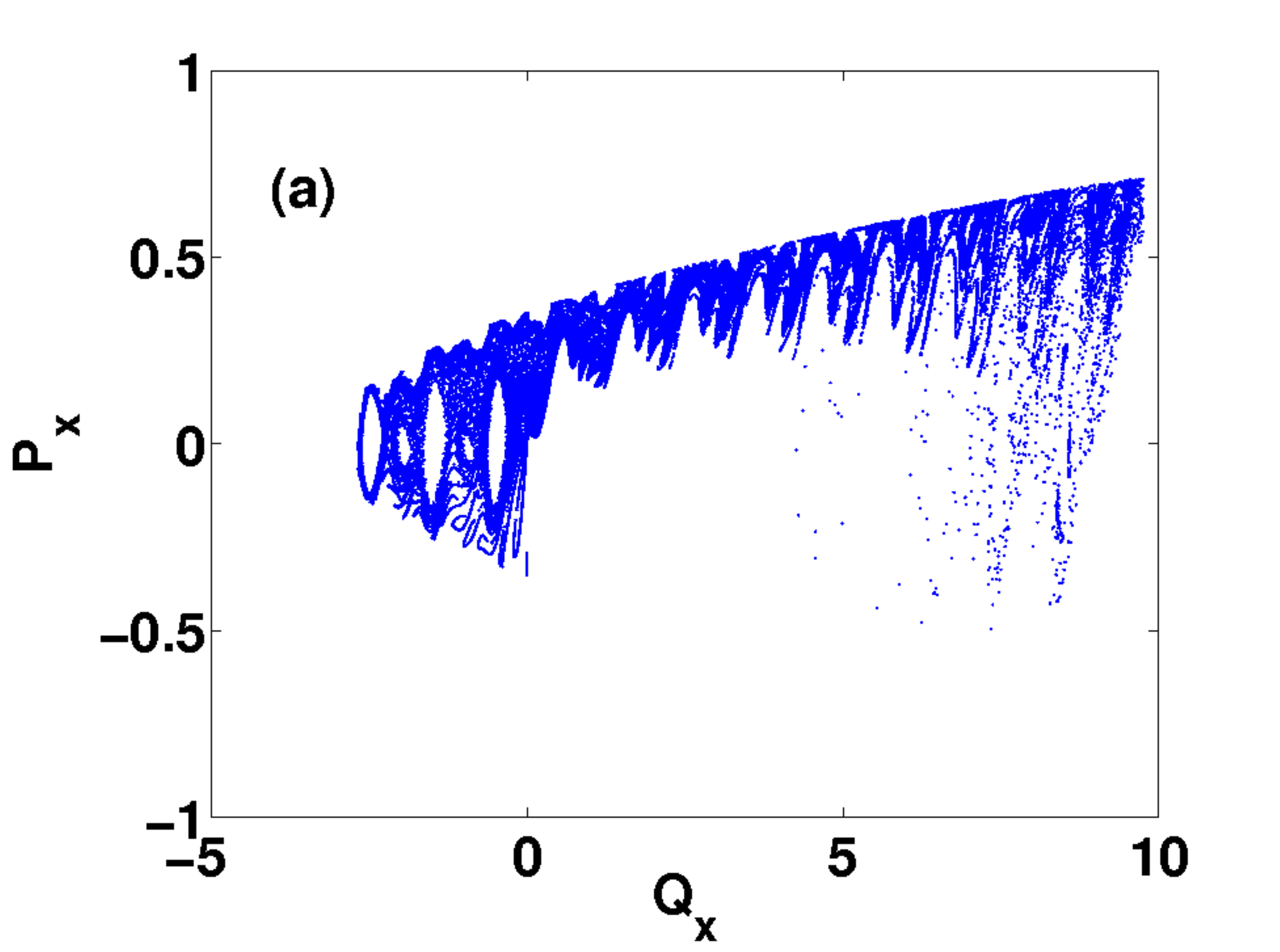}
\includegraphics[scale=0.3]{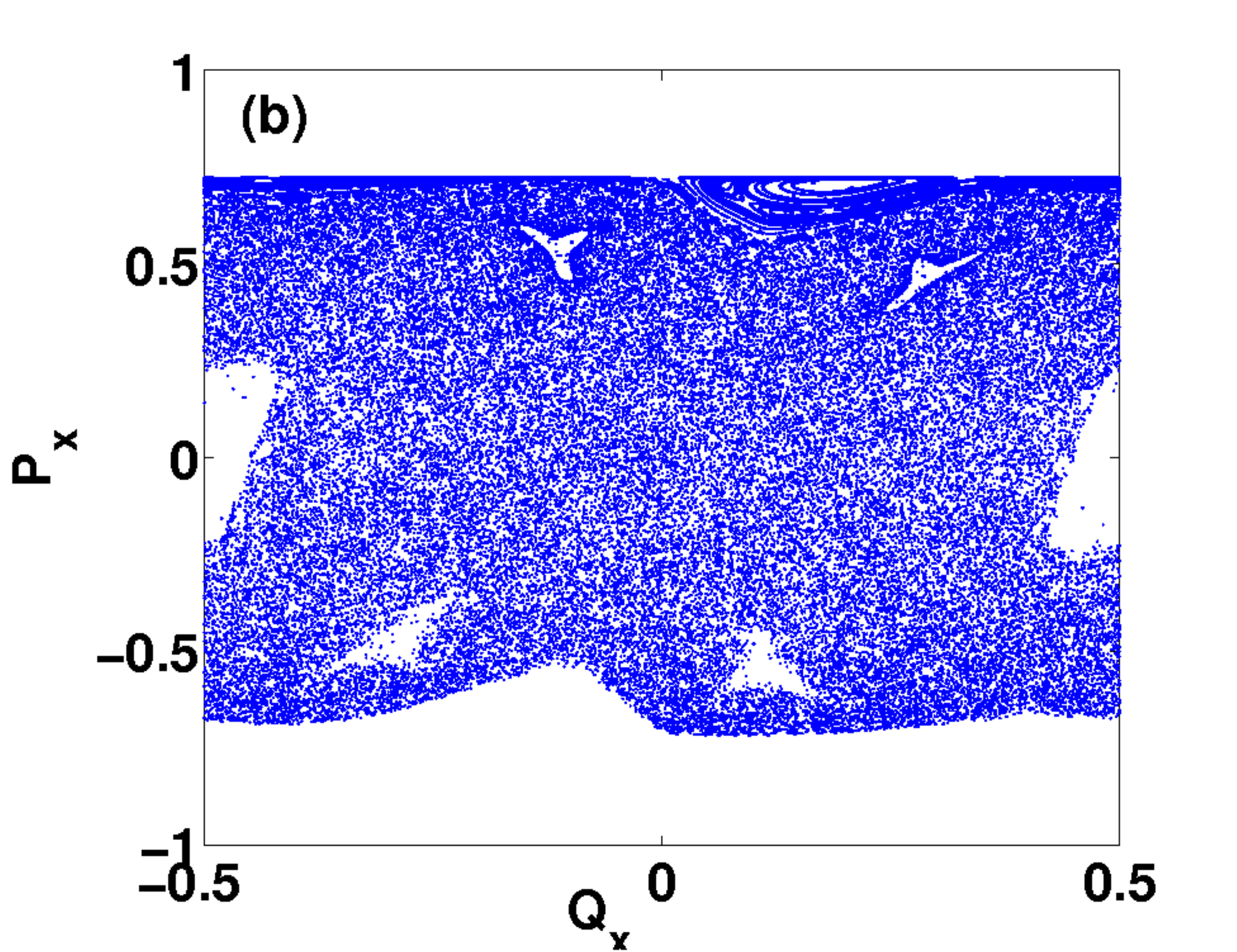}
\includegraphics[scale=0.3]{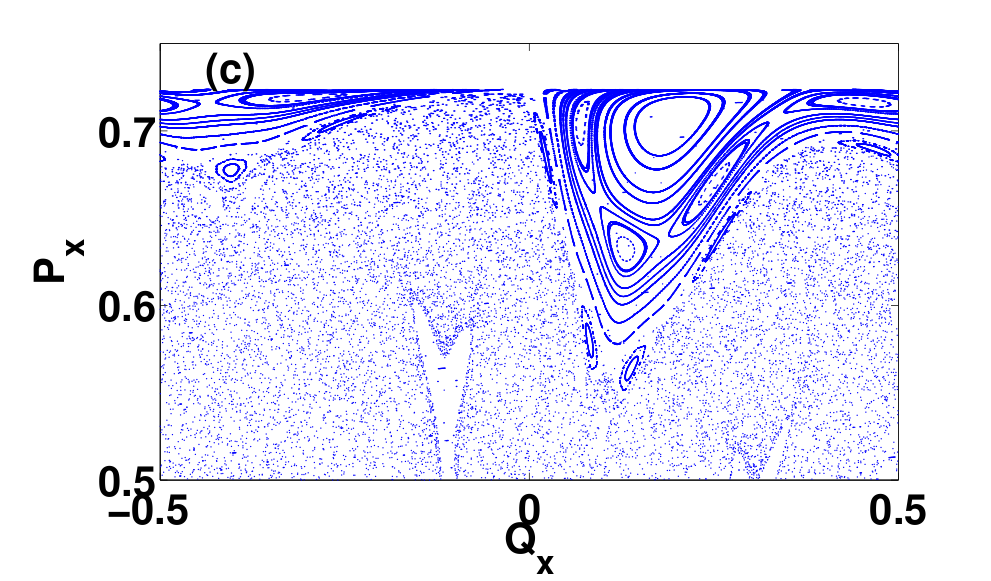}
\caption{(Colour online) PSS for the dynamics corresponding to a
  symmetrical uniform distribution of initial conditions satisfying
  relation~(\ref{eq:ring}) in the potential well at $Q_x=0$, showing
  (a) $Q_x\le 10$ and (b) the
  asymptotic regime $Q_x>10$ (shown $\mod(1)$), for coupling parameter
  $D=3$. The remaining parameter values are $\alpha=3$, $F=0.01$ and
  $q_0=10$.  The panel (c) shows detail of the mixed phase space at
  the upper boundary of the chaotic layer in (b).}
\label{fig:ballistic}
\end{figure}

Finally, we relate the escape process and the emergence of
directed chaotic motion to the phase space structure of the
transformed system with Hamiltonian given in Eq.~(\ref{equation:Hnew}).
To this end we use the following PSS
\begin{equation}
\Sigma=\left\{\,P_x, Q_x|Q_y=1/4, P_y>0\,\right\}\,.
\end{equation}
In Fig.~\ref{fig:ballistic}
we plot for a strong particle coupling $D=3$ the PSS corresponding to
the escape dynamics when $Q_x\le 10$ and the dynamics in the
asymptotic region, i.e. $Q_x>10$ (for those trajectories which reach
it), in (a) and (b) respectively, being characterised by chaotic sets.
(In Fig.~\ref{fig:ballistic}\,(b) the coordinate $Q_x$ is presented
mod$(1)$.) In Fig.~\ref{fig:ballistic}\,(a) chaotic saddles, formed by
the intersecting stable and unstable manifolds of unstable periodic
points, govern the dynamics. The majority of escaping trajectories
follows the unstable manifold of the saddle point located at
$(Q_x,Q_y)=(0.188,0.243)$. On the other hand there are trajectories
that remain in the starting region or spend at least some time there
before escape as a consequence of the presence of chaotic saddles
\cite{scattering},\cite{Alligood}.  Furthermore, on approaching the
asymptotic region, where the tilt of the washboard potential vanishes,
some of the previously-escaping dimers become trapped in wells of the
washboard potential again.

From Fig.~\ref{fig:ballistic}\,(b), depicting the PSS in the
asymptotic region, we conclude that the bulk of the layer on the PSS
is covered by a chaotic set. Within the chaotic set trajectories move
in a diffusive way with changes of the direction of motion not
contributing to transport.  Notably, at the upper boundary of the
layer, shown in Fig.~\ref{fig:ballistic}\,(c), islands of regular
motion arise from those trajectories that have settled on regular
dynamics after their passage through a chaotic transient.  Most
importantly, these islands possess non-zero winding numbers and thus
act as ballistic channels \cite{ballistic} providing directed
transport to the right.  In particular, the dynamics within the island
structure centered at the stable period-one fixed point
$(P_x,Q_x)=(0.703,0.180)$ reflects the almost-synchronous monomer
motion described in scenario (iv) in Section \ref{section:current}. In
more detail, motion near the fixed point corresponds to
almost-periodic energy exchange between the monomers
which is connected
with only minor bond deformations which corroborates the findings
reported above for the optimal transport scenario.

\section{Summary}

\noindent We have analysed the Hamiltonian dynamics of two nonlinearly
coupled particles evolving in a washboard potential.  Notably the
total energy does not suffice to enable simultaneous escape of the two
particles, initially trapped in a well of the washboard potential. Due
to appropriate energy redistribution, at least one of the particles
can achieve escape. (See also \cite{Simon}.) Ideally the two particles
share energy almost periodically in such a way that consecutive
detrapping-trapping transitions take place during which the particles
escape one after another from one well into an adjacent one. It has
been demonstrated that a weak tilt force, vanishing asymptotically in
the direction of the bias, is sufficient to instigate directed motion
of the escaping particles.  Transport is accomplished for those
trajectories which follow a chaotic transient, associated with the
dynamics of chaotic saddles, settling afterwards on regular motion.
The key mechanism of current rectification is based on the
asymptotically vanishing tilt causing a symmetry breaking of the
non-chaotic fraction of the dynamics where only at the upper boundary
of the chaotic layer resonance islands appear.  The latter is
supported by transporting island structures in the mixed phase space
which serve for long-range directed transport.

Finally, we mention that it is certainly interesting to extend the
study of directed motion to systems involving many more degrees of
freedom than in the current dimer case where the dynamics within
transporting islands can be investigated utilising two-dimensional
Poincar\'{e} surface of sections. In particular, it needs to be
examined what structures in higher dimensional phase spaces play the
role of possible ballistic channels providing directed collective
transport.

\end{document}